# Energetics of Cytoskeletal Gel Contraction

Matteo Ferraresso[1,2], Albert Kong[1,2], Mehadi Hasan[1,2], Daniele Agostinelli[1,2], Gwynn J. Elfring[1], Mattia Bacca[1,2,*]

[1]Mechanical Engineering Department, Institute of Applied Mathematics
[2]School of Biomedical Engineering
University of British Columbia, Vancouver BC V6T 1Z4, Canada

*Corresponding author: mbacca@mech.ubc.ca

**Abstract**

Cytoskeletal gels are prototyped to reproduce the mechanical contraction of the cytoskeleton in-vitro. They are composed of a polymer network (backbone), swollen by the presence of a liquid solvent, and active molecules (molecular motors, MMs) that transduce chemical energy into the mechanical work of contraction. These motors attach to the polymer chains to shorten them and/or act as dynamic crosslinks, thereby constraining the thermal fluctuations of the chains. We describe both mechanisms thermodynamically as a microstructural reconfiguration, where the backbone stiffens to motivate solvent (out)flow and accommodate contraction. Via simple steady-state energetic analysis, under the simplest case of isotropic deformation, we quantify the mechanical energy required to achieve contraction as a function of polymer chain density and molecular motor density. We identify two limit regimes, namely, fast MM activation (*FM*), and slow MM activation (*SM*). *FM* assumes that MMs provide all the available mechanical energy 'instantaneously' and leave the polymer in a *stiffened* state, *i.e.* the MM activity occurs at a time scale that is much smaller than that of solvent diffusion. *SM* assumes that the timescale for MM activation is much longer than that of solvent diffusion. To achieve the same final contracted state, *FM* requires the largest amount of work per unit reference volume, while *SM* requires the least. For all intermediate cases where the timescale of MM activation is comparable with that of solvent diffusion, the required work ranges between the two cases. We provide all these quantities as a function of chain density and MM density. Finally, we compare our results on contraction energetics with experiments and observe good agreement.

*Keywords*: Active soft matter; polymer gels; cytoskeleton; contraction; energetics

**Introduction**

The cytoskeleton is the structural backbone of the cell and is primarily made up of actin, intermediate filaments and microtubules [1]. Actin filaments coupled with myosin motors actively partake in the mechanical response of the cell and are responsible for its contractile behaviour. Actomyosin II complexes pull on the actin chains to shorten them, therefore the 3D meshwork of crosslinked actin chains contracts. This process is mainly powered by adenosine triphosphate (ATP) hydrolysis, where released chemical energy from the breakdown of ATP into ADP and a phosphate, is expended in the form of mechanical work. This is a critical function required for motility, environmental adaptation, chemical transport, and intracellular signaling [1].

Understanding cytoskeletal contraction from mechanical and energetic standpoints is critical to characterize tissue properties and tissue mechanics.

Cytoskeletal gels are engineered to mechanically model the cytoskeleton. They constitute a minimalistic physical system by including only the main components of the cytoskeleton: polymer chains and cytoplasm; while excluding organelles, membranes and other cell components [2,3]. Figure 1 outlines the main components of the gel and their arrangement in the system. These gels have been used to both characterize the mechanical properties of the cytoskeleton and to study cell contractility.

Multiple approaches to model the cytoskeleton and its active behaviours have been explored. An early bio-chemo-mechanical model for cytoskeletal contractility was introduced by Deshpande et al. [4]. Refined versions of this model were then developed to include thermodynamic motivations to cytoskeletal contractility [5] as well as experimental validation to explain the role of boundary conditions in defining the cell alignment mediated by the cytoskeleton [6]. Other early studies approached cytoskeletal modelling from a solid-mechanics foundation, with a focus on the effects of elasticity on cell shape and adhesion properties [7]. Gel models have also been widely adopted in modelling cytoskeletal mechanics; several models approached the problem from a hydrodynamic standpoint, where transient force dipoles, generated by myosin pulling on actin chains, create active contractile stresses [8,9]. The main limitation of such models lies in their assumption that actin filaments must not be slack, and this is not the case for some actin networks [9]. Furthermore, these models describe active stresses using parameters that are empirically determined, and typically lack physical meaning. Gels composed of slack chains are not well represented by these models as the cytoskeletal active stresses do not account for the level of tension in the chains. To overcome this limitation, recent studies propose an alternative approach based on poroelasticity, where the cytoskeleton is described as a polymer network swollen by the presence of the solvent (cytosol) [10-11]. In this biphasic material, the passive behavior is controlled by swelling and solvent flow [9,12-16], while the active behavior is described as an evolution of the mechanical stiffness of the polymer backbone [12] and is promoted by molecular motor activity. The total free energy of the gel is given by the sum of the strain energy of the network and the free energy (enthalpy and entropy) of chain-solvent mixing. The activation of contractile molecular motors (MM) increases the strain energy, representing a stiffening of the chain network [2,12,17-19]. This in turn increases the osmotic pressure of the solvent, locally, and, with it, the chemical potential of the solvent. Because the flow of the solvent is directed toward a reduction in chemical potential, via Fick's law, the solvent ultimately outflows from the stiffened gel, thereby accommodating gel contraction [12,20-21]. Polymer network stiffening can be achieved via dynamic crosslinking (DC) created by a MM attaching to more than one chain. This is described in some models as an evolution of the crosslink density [12,20].

The observed MM activity in actin cytoskeletal gels primarily involves chain shortening (CS) rather than DC [2]. Figure 1 presents the two motor activation mechanisms. An energetically based model incorporating realistic MM activation applicable to in-vivo environments is therefore necessary to fully understand actin gel contractile mechanics. This study builds on active polymer gel theory and introduces a new molecular motor model considering chain shortening (CS-MM), striving towards a more accurate representation of the physical system. Our model introduces a new thermodynamically inspired variation of neo-Hookean chain strain energy to incorporate a CS parameter we define as the 'microstretch'. This parameter decouples chain density from contractile mechanics, allowing for direct comparison between the energetic requirement for a specific contraction (quantity of molecular motors) and initial chain density.

Experiments using an *in-vitro* actin network platform reveal a threshold for the *minimum* and *maximum* densities of MMs required to achieve a desired contraction, based on the crosslink density of the gel [22-23]. Via quasi-static energetic analysis, we use our model to predict this threshold in a direct comparison with experimental data of contractile acto-myosin networks. In this comparison, we introduce two contraction regimes: *slow* MM activation (S*M*), where the time scale for MM activation is longer than that for solvent outflow, and *fast* MM activation (*FM*), where contraction is triggered by an instantaneous injection of all MM work available (*i.e.* the timescale for motor activation is shorter than solvent outflow). The work performed by MMs is higher in the *FM* regime than in the *SM* one. Thus, we take the energetic requirement from the *SM* regime to predict the *minimum* MM density required for a desired contraction. To predict the *maximum* MM density, for a given crosslink density, we compute the maximum strain energy accumulated in each polymer chain prior to chain rupture [24-25]. Finally, our predictions for *minimum* and *maximum* MM density are in good agreement with the experiments.

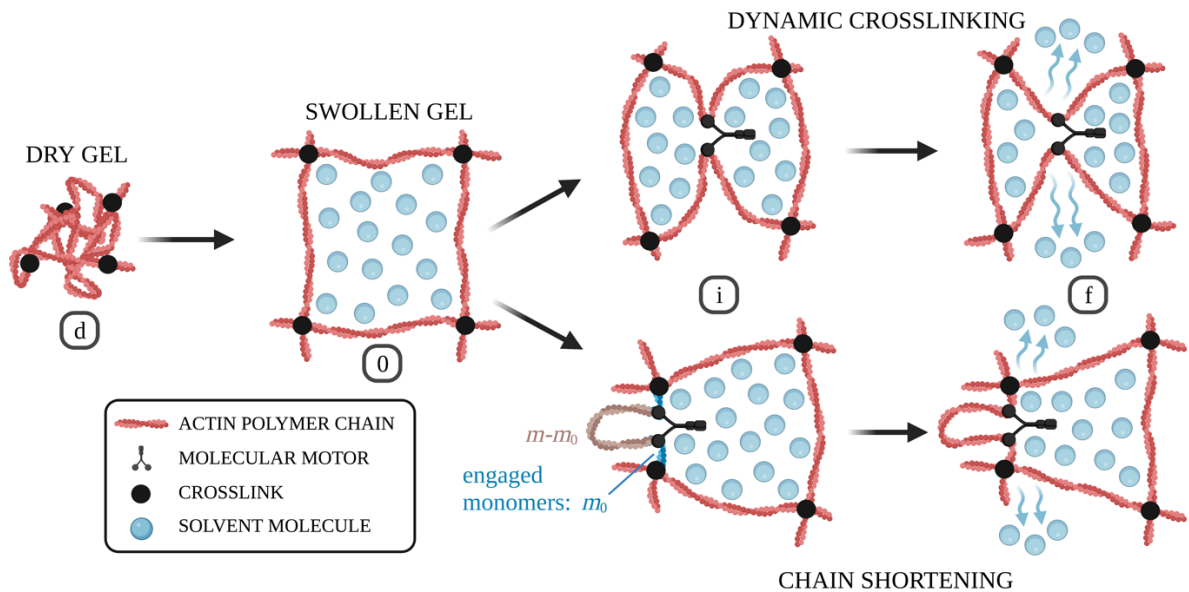

**Figure 1** – Diagram outlining the two main mechanisms of cytoskeletal gel contraction from molecular motor activation: dynamic crosslinking molecular motors (DC-MM, top) and chain shortening molecular motors (CS-MM, bottom). From left to right, the gel is initially in its reference dry configuration (d). It is then swollen by solvent (blue spheres) and achieves chemo-mechanical equilibrium with the environment (0). When activated, the motors inject energy by stiffening the swollen polymer backbone (i), which causes solvent outflow accommodating contraction (f). DC-MM (top) attach different chains in the network, acting as a temporary crosslink, thereby increasing the crosslink density, whereas CS-MM (bottom) decrease the average polymer chain length. The latter reduces the number of engaged monomers, from $m$ to $m_0 < m$.

**Model: Active Gel**

The biphasic active gel is composed of a (swollen) polymer network, a solvent, and molecular motors (MM) (Fig. 1). The polymer chains provide the structural backbone of the gel, and are stretched to accommodate the presence of the solvent. We define the dry polymer network (prior to solvent absorption and swelling) as the stress-free reference state of the gel (albeit for some gels this is an idealized state since the polymer dissociates in the absence of solvent).

The total free energy density, per unit *reference volume* $V_d$, as described by [12], is

$$\psi = \psi_e(T, \boldsymbol{F}, \boldsymbol{\zeta}) + \psi_m(T, C) + \Pi(1 + \Omega C - J) \tag{1}$$

where $\psi_e$ is the elastic strain energy density of the polymer network, $\psi_m$ is the free energy density of polymer-solvent mixing, $\Pi$ is the hydrostatic pressure applied to the gel, $C$ is the molecular concentration of solvent, per unit reference volume, $\Omega$ is the molecular volume of the solvent and $J$ is the swelling ratio of the gel (current volume to reference volume). In Eq. (1), $T$ is the absolute temperature, $\boldsymbol{F}$ is the deformation tensor defined as

$$F_{jk} = \frac{\partial x_j}{\partial X_k} \tag{2}$$

With $\boldsymbol{x}$ the position of the material points in the current state and $\boldsymbol{X}$ that in the reference state, $\boldsymbol{\zeta}$ is the collection of microstructural parameters that govern the elastic response of the polymer network such as polymer chain density and chain length. $\Pi$ serves as a Lagrange multiplier to enforce molecular incompressibility so that, with respect to the reference configuration, the swelling ratio $J$ is defined as

$$J = 1 + \Omega C \tag{3}$$

Where, in the reference configuration, $J = 1$ when no solvent is present (only polymer) and increases with increasing solvent concentration $C$ in proportion to the molecular volume of the solvent $\Omega$. Considering the solvent bath is isothermal, from the first and second laws of thermodynamics [8] (see Appendix A), we obtain the relationship between stresses, chemical potential and the Helmholtz free energy from Eq. (1) as follows

$$t_{jk} = \frac{\partial \psi}{\partial F_{jk}} \tag{4}$$

$$\mu = \frac{\partial \psi}{\partial C} \tag{5}$$

and

$$dw_c = \frac{\partial \psi}{\partial \zeta_c} d\zeta_c \tag{6}$$

where $\boldsymbol{t}$ is the first Piola-Kirchoff stress tensor, $\mu$ is the chemical potential of the solvent, $dw_c$ is the unit work, per unit reference volume, required to produce a unit change $d\zeta_c$ in the microstructural variable $\zeta_c$. This work is here provided by the molecular motors as part of the energy consumed by ATP hydrolysis.

To ensure mechanical equilibrium, we need

$$\frac{\partial t_{jk}}{\partial X_k} + B_j = 0, \text{ in } V_d \tag{7a}$$

and

$$t_{jk}N_k = T_j, \quad \text{on } S_d \tag{7b}$$

with $\boldsymbol{B}$ the body force vector, $\boldsymbol{N}$ the unit outward normal to $S_d$, i.e. the boundary of $V_d$, and $\boldsymbol{T}$ the external traction force. $\boldsymbol{B}$ is a force per unit $V_d$, while $\boldsymbol{T}$ is a force per unit $S_d$.

Assuming the gel boundary $S_d$ is permeable, a free exchange of solvent molecules with the external bath imposes

$$\mu = \mu_{ext} \text{ on } S_d \tag{8a}$$

with $\mu_{ext}$ the chemical potential of the solvent in the bath. Conversely, in the case of impermeable boundary, one should instead impose

$$H_i N_i = 0 \text{ on } S_d \tag{8b}$$

(with repeated indices indicating the sum) with $\boldsymbol{H}$ the solvent flux given by Fick's law [6-7] as

$$H_i = -DC F_{ik}^{-1} F_{jk}^{-1} \frac{\partial(\mu/kT)}{\partial X_j} \tag{8c}$$

with $D$ the diffusivity of the solvent, and $k$ Boltzmann's constant.

Eq. (7b) gives the mechanical boundary conditions to Eq. (7a), with substitution of Eq. (1) and (4) into it. Eq. (8a-b) instead give the chemical boundary conditions to Eq. (8c), together with mass conservation (see Appendix A), with substitution of Eq. (1) and (5) into it.

Chemical equilibrium imposes $\boldsymbol{H} = \boldsymbol{0}$, i.e. no solvent flow, and from Eq. (8c) we can deduce that this condition imposes a spatially homogeneous chemical potential $\mu$. Assuming at least a portion of $S_d$ is permeable, we have that chemical equilibrium imposes

$$\mu = \mu_{ext} \text{ in } V_d \tag{9}$$

*Mixing free energy*

Assuming the polymer network is diluted into the gel (polymer-solvent mixture), i.e. $J \gg 1$, $\psi_m$ can be described with the bimolecular mixing free energy theory introduced by [26] and giving

$$\frac{\Omega \psi_m}{kT} = \Omega C \left[ \ln\left(\frac{\Omega C}{1+\Omega C}\right) + \frac{\chi}{1+\Omega C} \right] \tag{10}$$

The first term on the right-hand side of Eq. (10) provides the entropic contribution to the mixing free energy, and provides the configurational energy reduction (favoring mixing) of solvent molecules relocating from the solvent bath to the gel. The second term describes the energy associated with steric interactions between solvent and polymer, where $\chi$ is the Flory parameter [26]. $\chi > 0$ describes repulsive interactions (favoring demixing) as, in this case, the free energy in Eq. (10) increases with $J$. Conversely, $\chi < 0$ describes attractive interactions (favouring mixing) as, in this case, the free energy in Eq. (10) reduces as $J$ increases.

*Elastic free energy*

In our model the polymer backbone is composed of a network of Gaussian chains with variable length, yielding a modified neo-Hookean elastic free energy density (see Appendix B)

$$\frac{\Omega \psi_e}{kT} = \frac{n}{2}\left[F_{jk}F_{jk}\lambda_M^{-2} - 2\ln(J\lambda_M^{-3}) - 3\right] \tag{11a}$$

with

$$n = N\Omega \tag{11b}$$

Here, $N$ ($n$) is the chain density (dimensionless chain density) per unit reference volume (unit solvent molecule) and $\lambda_M = m/m_0 \leq 1$ is a chain micro-stretch parameter that incorporates the effect of chain shortening in the elastic free energy, where $m$ is the average number of *engaged* monomers connecting the two ends of the chain, and $m_0$ is the total number of monomers in the chain (see Figure 1). In the case of $\lambda_M = 1$ no chain microstretch occurs and the strain energy expression becomes that of a swollen incompressible neo-Hookean material. In the case of $\lambda_M < 1$, we have $m_0 < m$ engaged monomers, so that we have $m - m_0$ monomers that are excluded from the description of the configurational energy of the chain. Here, the two independent microstructural variables ($\zeta$) being considered are $n$ and $\lambda_M$. Dynamic crosslinking molecular motors (DC-MM) are associated with an evolution of chain density $n$, as further polymerization with additional crosslinker molecules increases the number of active chains in the gel. Chain-shortening molecular motors (CS-MM) are associated with an evolution of the microstretch $\lambda_M$, where the average effective chain length reduces, *i.e.* the effective number of monomers connecting the two ends of the chains reduces. Here, the polymer fraction does not evolve as we assume the newly 'unengaged' monomers remain within the gel. Thus, mass conservation enforced by Eq. (3) applies at all times. We analyze here DC-MM and CS-MM contraction mechanisms separately. Additionally, note that $N$ is defined in the reference (dry polymer) state, while the chain density in the current state is $\widehat{N} = N/J$.

*Isotropic contraction of unloaded gels*

Consider now an unloaded gel. By substituting Eq. (11) into (1), and the result into (4), the Cauchy stress $\sigma_{ij} = t_{ik}F_{jk}/J$ takes the form

$$\frac{\Omega \sigma_{jk}}{kT} = \frac{n}{J}\left(F_{ji}F_{ki}\lambda_M^{-2} - \delta_{jk}\right) - \frac{\Omega \Pi}{kT}\delta_{jk} \tag{12}$$

In the absence of body forces, we can assume the stress components are spatially homogeneous within the gel. Because the stress has to equilibrate the external pressure $\Pi_{ext}$ applied to the gel from the environment (the solvent bath), we have that $\sigma_{jk} = -\Pi_{ext}\delta_{jk}$ for each $j,k$ couple. By imposing this condition to Eq. (12), we must have $F_{ji}F_{ki} = \delta_{jk}$. This condition is satisfied in the case of isotropic swelling of the gel, where the deformation is characterized by equal principal stretches $\lambda$ in all directions as function of the swelling ratio $J = \lambda^3$. Substituting this condition into Eq. (12) we have the hydrostatic gel pressure

$$\Pi = \Pi_{ext} + \Pi_{osm} \tag{13a}$$

where

$$\frac{\Omega \Pi_{osm}}{kT} = \frac{n}{J}\left(J^{2/3}\lambda_M^{-2} - 1\right) \tag{13b}$$

is the osmotic pressure of the solvent, which is also the hydrostatic swelling stress that stretches the polymer chains to accommodate the presence of the solvent.

By substituting Eq. (10) into (1), and the result into (5), we obtain the chemical potential of the solvent as

$$\frac{\mu}{kT} = \ln\left(\frac{\Omega C}{1+\Omega C}\right) + \frac{1}{1+\Omega C} + \frac{\chi}{(1+\Omega C)^2} + \frac{\Omega \Pi}{kT} \tag{14}$$

Substituting Eq. (3) and (13) into (14), we have

$$\bar{\mu} = \ln\left(1 - \frac{1}{J}\right) + \frac{1}{J} + \frac{\chi}{J^2} + \frac{n}{J}\left(J^{2/3}\lambda_M^{-2} - 1\right) \tag{15a}$$

where

$$\bar{\mu} = \frac{\mu - \Omega \Pi_{ext}}{kT} \tag{15b}$$

is the dimensionless net chemical potential. The gradient of chemical potential, and thus that of $\bar{\mu}$, drives the solvent flow from substitution of Eq. (15b) into (8c). The first two terms on the right-hand side of Eq. (15a) provide the entropic contribution, the third term provides the enthalpic contribution, and the last term, multiplied by $n$, provides the elastic contribution from the polymer network. At chemo-mechanical equilibrium we have $\bar{\mu} = \bar{\mu}_{ext}$, with $\bar{\mu}_{ext}$ obtained from Eq. (15b) by substituting $\mu$ with $\mu_{ext}$.

From Eq. (15) we can obtain the derivatives of $\bar{\mu}$ with respect to the dimensionless microstructural variables $n$ and $\lambda_m$ as

$$\frac{\partial \bar{\mu}}{\partial n} = \frac{1}{\lambda_M^2 J^{1/3}} - \frac{1}{J} \tag{16a}$$

$$\frac{\partial \bar{\mu}}{\partial \lambda_M} = -\frac{2n}{\lambda_M^3 J^{1/3}} \tag{16b}$$

Eq. (16) shows that polymer stiffening via increment of $n$ (DC-MM) or reduction of $\lambda_M$ (CS-MM) always produces an increment of $\mu$. So if $\bar{\mu} = \bar{\mu}_{ext}$, before any MM activity, we will have $\bar{\mu} > \bar{\mu}_{ext}$ after MMs are activated. According to Eq. (8c), with substitution from Eq. (15b), the solvent migrates from high-$\bar{\mu}$ regions to low-$\bar{\mu}$ ones. Thus, in this case, the solvent will exit the gel to join the buffer (external solvent bath). This process reduces $\bar{\mu}$ to recover chemical equilibrium. This occurs if

$$\frac{1}{\Omega}\frac{\partial \bar{\mu}}{\partial C} = \frac{1}{J^2(J-1)} - \frac{2\chi}{J^3} + \frac{n}{3J^2}\left(3 - J^{2/3}\lambda_M^{-2}\right) \tag{17}$$

is positive so that $\bar{\mu}$ decreases to $\bar{\mu}_{ext}$ while $J$ decreases to its new equilibrium value. The positivity of the right-hand side of Eq. (17) depends on $\chi$, $N$, $\lambda_M$ and $J$, thus posing some physical constraints on these variables.

Contraction spontaneity requires free energy reduction at a higher rate than that of energy supplied by MMs, i.e. $\dot{\psi} \leq \dot{w}$, where $\dot{w}$ is the power produced by the MMs. This gives

$$\frac{\partial \psi_e}{\partial J}\dot{J} + \frac{\partial \psi_m}{\partial C}\dot{C} + \frac{\partial \psi_e}{\partial N}\dot{N} + \frac{\partial \psi_e}{\partial \lambda_M}\dot{\lambda}_M \leq \dot{w} \tag{18a}$$

From Appendix A and Eq. (6), we have $\dot{w} = \dot{N}\,\partial\psi_e/\partial N + \dot{\lambda}_M\,\partial\psi_e/\partial\lambda_M$, from which we can rewrite Eq. (18a) as

$$\frac{\partial \psi_e}{\partial J}\dot{J} + \frac{\partial \psi_m}{\partial C}\dot{C} \leq 0 \tag{18b}$$

Here, $\partial \psi_e / \partial J = -\Pi_{ext}$ and, by substituting Eq. (3), (5) and (15) into (18), we finally have the following condition for the spontaneity of contraction ($\dot{C} < 0$) of swelling ($\dot{C} > 0$)

$$\bar{\mu}\dot{C} \leq 0 \tag{19}$$

Spontaneous contraction, $\dot{C} < 0$, requires $\bar{\mu} \geq 0$, while swelling, $\dot{C} > 0$, requires $\bar{\mu} \leq 0$. When $\bar{\mu} \neq 0$ the rate of free energy release is finite (under contraction or swelling) leading to finite contraction or swelling times. When $\bar{\mu} = 0$ the energy release rate is equal to zero leading to infinite contraction or swelling times. This provides the condition of chemo-mechanical equilibrium, and is in agreement with the condition $\bar{\mu} = \bar{\mu}_{ext}$, where $\bar{\mu}_{ext} = 0$. In the solvent bath, $\mu_{ext} = \Omega(P - P_{vap})$ [10], with $P$ the solvent pressure and $P_{vap}$ the vapor (cavitation) pressure. Taking $\Pi_{ext} = P - P_{vap}$ as the relative pressure of the solvent bath, we finally have $\bar{\mu}_{ext} = 0$.

From Eq. (3) and (15), we can define the dimensionless chemical potential $\bar{\mu}$ as

$$\bar{\mu} = \frac{1}{\Omega}\frac{d\bar{\psi}}{dC} \tag{20a}$$

with

$$\frac{d\bar{\psi}}{dC} = \frac{\partial \bar{\psi}_m}{\partial C} + \Omega \frac{\partial \bar{\psi}_e}{\partial J} \tag{20b}$$

where $\bar{\psi} = \psi \Omega/kT$, $\bar{\psi}_m = \psi_m \Omega/kT$, and $\bar{\psi}_e = \psi_e \Omega/kT$ are dimensionless forms of free energy. From this, we can deduce that the positivity of the left-hand side of Eq. (17), under chemo-mechanical equilibrium ($\bar{\mu} = 0$), occurs at a free energy minimum, and therefore such a chemo-mechanical equilibrium is stable.

In the following derivation, we will compare the free energy at three states: (0) the swollen equilibrium state, prior to MM activation; (*i*) the initial state when MM are activated and provide all the available mechanical work to stiffen the backbone; (*f*) the final state at which all the contraction has occurred and the stiffened gel has reached a new equilibrium configuration. The swelling ratio in (0) and (*i*) is the same, assuming MM activity is much faster than solvent diffusion. In this case $J = J_0$.

For the generic state $\varrho = 0, i, f$, the dimensionless free energy can be rewritten from Eq. (1), (10) and (11) as

$$\bar{\psi}_\varrho = \bar{\psi}_{e,\varrho} + \bar{\psi}_{m,\varrho} \tag{21a}$$

$$\bar{\psi}_{e,\varrho} = \frac{n_\varrho}{2}\left[3(J_\varrho^{2/3}\lambda_{M,\varrho}^{-2} - 1) - 2\ln(J_\varrho\lambda_{M,\varrho}^{-3})\right] \tag{21b}$$

$$\bar{\psi}_{m,\varrho} = (J_\varrho - 1)\ln\left(1 - \frac{1}{J_\varrho}\right) + \chi\left(1 - \frac{1}{J_\varrho}\right) \tag{21c}$$

where $n_\varrho = N_\varrho \Omega$ is the dimensionless crosslink density, $\lambda_{M,\varrho}$ the chain micro-stretch, and $J_\varrho$ the swelling ratio at the state $\varrho = 0, i, f$.

In the swollen equilibrium state (0), $J_0$ is determined from Eq. (15) at chemical equilibrium by imposing the condition $\bar{\mu} = 0$ in Eq. (15a), with $n = n_0$, $\lambda_M = 1$, and $J = J_0$.

To achieve contraction, the swelling ratio must reduce from the initial state (*i*) to the final (*f*), thus $J_f < J_i$. At the initial state (*i*), while $J_i = J_0$, the microstructure has evolved via an increase in chain density, giving $n_i = N_i\Omega > n_0$ (DC-MM), and/or shortening of chains, giving $\lambda_{M,i} < 1$ (CS-MM).

Both these mechanisms induce network stiffening, thereby increasing the osmotic pressure, which in its dimensionless form $\bar{\Pi}_{osm} = \Pi_{osm}\Omega/kT$, from Eq. (13b), becomes

$$\bar{\Pi}_{osm,i} = \frac{n_i}{J_0}\left(\frac{J_0^{2/3}}{\lambda_{M,i}^2} - 1\right) \tag{22}$$

Now the free energy is given by Eq. (21) with $J_i = J_0$, $n = n_i$, and $\lambda_M = \lambda_{M,i}$. From the equilibrium state (0) to the initial one (i), the mixing energy in unchanged giving $\bar{\psi}_{m,i} = \bar{\psi}_{m,0}$, while the elastic strain energy has increased to give $\bar{\psi}_{e,i} > \bar{\psi}_{e,0}$, and thus $\bar{\psi}_i > \bar{\psi}_0$.

To define the final state (f) we must impose again chemical equilibrium. This involves the condition $\bar{\mu} = 0$ in Eq. (15a) with $n = n_i$, $\lambda_M = \lambda_{M,i}$, and $J = J_f$. The free energy in this state is given again by Eq. (21) with $J_f$, $n_f = n_i$, and $\lambda_{M,f} = \lambda_{M,i}$.

The total change in volume, from the swollen equilibrium state (0) to the final contracted state (f), is given by the ratio $dV_f/dV_0 = J_f/J_0$, with $dV_0$ and $dV_f$ the unit volumes in the current (0) and (f) states. The ratio $J_f/J_0$ identifies the degree of contraction, with smaller $J_f/J_0$ requiring higher contraction. Higher contraction is achieved by increasing $n_i/n_0$ and/or $1/\lambda_{M,i}$, and, as shown in the Results and Discussion section, the contraction amount depends also on the initial conditions defined by $J_0$ and $n_0$.

We now introduce the microstructural mechanical work, $w_M$, per unit reference volume, (in dimensionless form, $\bar{w}_M = w_M \Omega/kT$) required to create contraction and performed by the MM. We introduce two extreme cases of MM activated contraction: *fast* MM activation (*FM*), where a sudden injection of all the microstructural work available triggers contraction and *slow* MM activation (*SM*), where the time scale for MM activation is much longer than that for solvent flow. The work due to *fast* MM activation (*FM*) is

$$\bar{w}_{FM} = \bar{\psi}_i - \bar{\psi}_0 \tag{23}$$

In this case, we assume that the time scale for MM activation is much shorter than that due to solvent flow. Additionally, we assume that, after the MMs perform all the microstructural work to stiffen the polymer, they stay in place keeping the polymer in its new stiffened state. This means that $\lambda_M$ and $n$ are stationary from (i) to (f), i.e. $\lambda_{M,f} = \lambda_{M,i}$ and $n_i = n_f$. Conversely, in the case of *slow* MM activation (*SM*), the work performed by the MM to achieve the final state is

$$\bar{w}_{SM} = \bar{\psi}_f - \bar{\psi}_0 \tag{24}$$

Because any contraction is spontaneous, we will always have a lower free energy in the final state than in the initial, thus $\bar{\psi}_i \geq \bar{\psi}_f$ and $\bar{w}_{FM} \geq \bar{w}_{SM}$. In intermediate cases, for which the time scale for MM activation is comparable with that of solvent flow, we have $\bar{w}_{SM} \leq \bar{w}_M \leq \bar{w}_{FM}$. This means that slow activation requires the minimum work $\bar{w}_{SM}$ to reach the state *f*, while fast activation requires the maximum work $\bar{w}_{FM}$ for the same purpose.

**Results and discussion**

*Chemo-mechanical Equilibrium*

As discussed in the previous section, chemo-mechanical equilibrium is obtained from the condition of a homogeneous distribution of chemical potential, equal to that of the solvent bath, thus, $\bar{\mu} = 0$. From Eq. (15), this imposes the following condition on the crosslink density

$$n = \frac{1 + \chi/J + J \ln(1 - 1/J)}{1 - J^{2/3} \lambda_M^{-2}} \tag{25a}$$

Given a microstretch $\lambda_M$ and Flory parameter $\chi$, the swelling ratio $J$ and (dimensionless) chain density $n$ at equilibrium are coupled according to Eq. (25). From this equation, Figure 2-left reports the values of compatible $J$ and $n$ for different values of $\chi$ and $\lambda_M$. In this figure, one can observe that a larger (dimensionless) crosslink density $n$ produces a smaller equilibrium swelling ratio $J$, as a stiffer gel can accommodate less volumetric swelling. The log-log plot in this figure shows the trend

$$n \sim \lambda_M^2 \frac{1 - 2\chi}{2} J^{-5/3} \tag{25b}$$

for large $J$ and $\chi < 0.5$, while for $\chi = 0.5$ we have $n \sim \lambda_M^2 J^{-8/3}/3$. Taking the microstretch as $\lambda_M = 1$, from Figure 2 we can obtain the equilibrium swelling ratio of the passive gel $J_0$ from the initial crosslink density $n_0$. This defines the initial state, prior to molecular motor (MM) activation, and is marked by blue circles on Figure 2. In the experimental setup used for model comparison [23], the measured values are $J_0 = 1000$, $n_0 = 1.61 \cdot 10^{-5}$, $\lambda_{M,0} = 1$, thus, yielding $\chi = -1.1$ (see Appendix C). In the case of dynamic crosslinking molecular motors (DC-MM), from the same figure, given the microstretch remains $\lambda_M = 1$, we can also determine the final swelling ratio $J_f$ associated with a final crosslink density $n_f$. For all the datapoints in Figure 2-left, the positivity of

$$\frac{1}{\Omega} \frac{\partial \bar{\mu}}{\partial C} = \frac{1}{J^2(J-1)} - \frac{2\chi}{J^3} + \frac{J \ln(1 - 1/J) + 1 + \chi/J}{1 - J^{2/3} \lambda_M^{-2}} \left( \frac{1}{J^2} - \frac{1}{3 J^{4/3} \lambda_M^2} \right) \tag{26}$$

from Eq. (17), is always satisfied. *I.e.* the chemo-mechanical equilibrium at $\bar{\mu} = 0$ is always stable.

By rearranging Eq. (25), one can also obtain

$$\lambda_M^2 = \frac{n J^{2/3}}{n - 1 - \chi/J - J \ln(1 - 1/J)} \tag{27a}$$

Figure 2-right reports the values of compatible swelling ratio $J$ versus microstretch $\lambda_M^2$ for different values of $\chi$ and $n$, based on Eq. (27). Because a smaller microstretch $\lambda_M^2$ yields a stiffer gel, and thus smaller equilibrium swelling ratio $J$, $\lambda_M^2$ and $J$ are proportional. The log-log plot in this figure shows the trend

$$\lambda_M^2 \sim \frac{2n}{1 - 2\chi} J^{5/3} \tag{27b}$$

for large $J$ and $\chi < 0.5$ (which can also be extrapolated from Eq. (25b)), while for $\chi = 0.5$ we have $\lambda_M^2 \sim 3n J^{8/3}$. In the case of chain-shortening molecular motors (CS-MM), this figure can be used to identify $J_f$ from a given $n = n_0$ and $\lambda_{Mf} < \lambda_{Mi}$. For all the datapoints in Figure 2-right, the positivity of

$$\frac{1}{\Omega} \frac{\partial \bar{\mu}}{\partial C} = \frac{2 + J}{3 J^2 (J-1)} + \frac{2}{3} \frac{n}{J^2} - \frac{5}{3} \frac{\chi}{J^3} + \frac{1}{3J} \ln\left(1 - \frac{1}{J}\right) \tag{28}$$

from Eq. (17), is again always satisfied. Thus, the chemo-mechanical equilibrium is again always stable.

From Eq. (25b) and (27b), under the assumption of large swelling ratio $J$ and $\chi < 0.5$, we can write

$$J \sim \left(\frac{1/2 - \chi}{n \lambda_M^{-2}}\right)^{3/5} \tag{29}$$

while for $\chi = 0.5$ we have $J \sim (3n\lambda_M^{-2})^{-3/8}$. In Eq. (29) we can deduce that largely swollen gels are made of softer polymers, having smaller $n \lambda_M^{-2}$, and/or polymers composed of chains that have high affinity with solvent molecules, thus smaller (more negative) $\chi$.

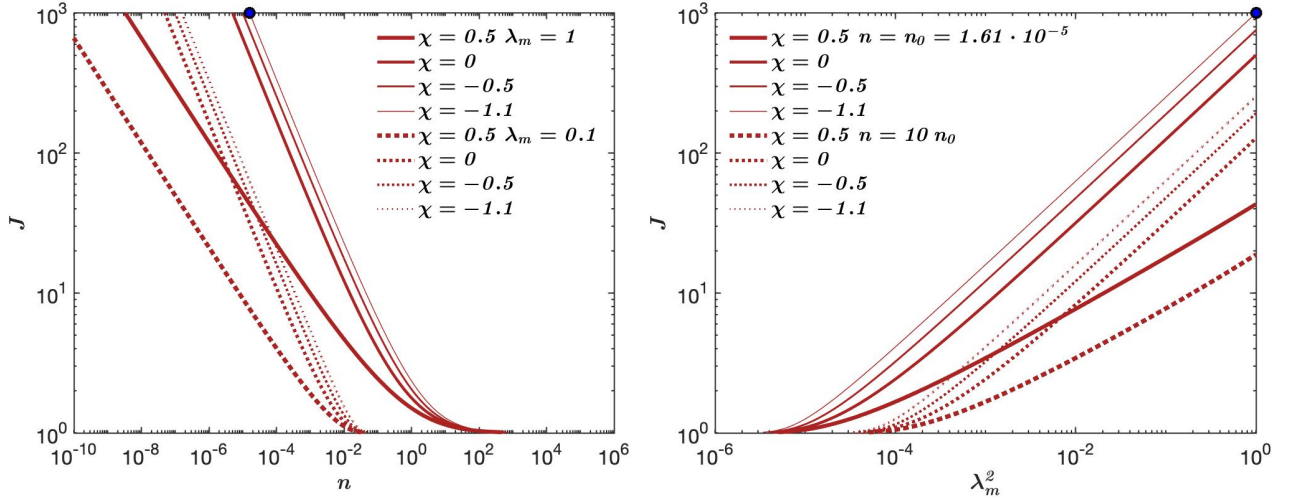

**Figure 2** – Cytoskeletal gel chemo-mechanical equilibrium conditions for dynamic crosslinking molecular motors (DC-MM) (left) and chain shortening molecular motors (CS-MM) (right). *Left*: From Eq. (25), network chain density per unit solvent molecule $n = N\Omega$ versus equilibrium swelling ratio $J$ at various values of the Flory parameter $\chi$ and chain microstretch $\lambda_M$. *Right*: From Eq. (27), chain microstretch $\lambda_M^2 = m/m_0$, as ratio of engaged monomers to total monomers in a chain, versus equilibrium swelling ratio $J$, at various values of the Flory parameter $\chi$ and chain density $n$. Blue circles represent the equilibrium state of the passive gel from experimentally measured parameters [23] (see Appendix C).

Let us now define the free energy from Eq. (21). Figure 3 (left) plots $\bar{\psi}_e/n$ versus $J$ and $\lambda_M$, and (right) $\bar{\psi}_m$ versus $J$ and $\chi$. These are taken from

$$\frac{\bar{\psi}_e}{n} = \frac{3}{2}\left(J^{2/3}\lambda_M^{-2} - 1\right) - \ln(J\lambda_M^{-3}) \tag{30a}$$

$$\bar{\psi}_m = (J - 1)\ln\left(1 - \frac{1}{J}\right) + \chi\left(1 - \frac{1}{J}\right) \tag{31a}$$

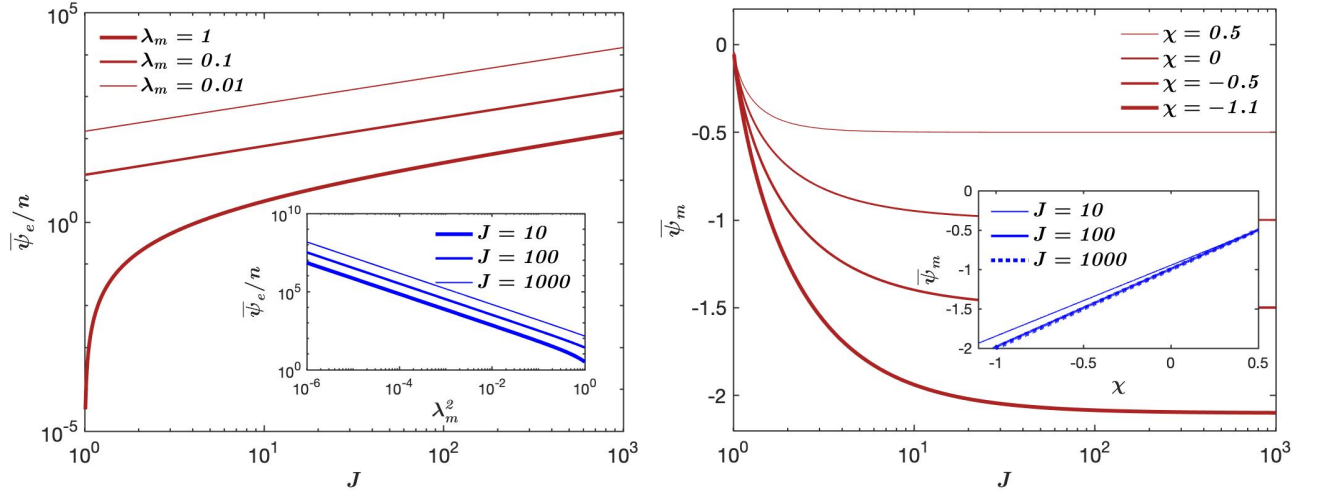

**Figure 3** – *Left*: Dimensionless elastic free energy ratio $\bar{\psi}_e/n$ versus gel swelling ratio $J$ at varying chain microstretch $\lambda_M$ (line thickness), from Eq. (30a); *Subplot*: $\bar{\psi}_e/n$ versus $\lambda_M$ at varying $J$ (line thickness), from Eq. (30a). *Right*: Dimensionless mixing energy $\bar{\psi}_m$ versus swelling ratio $J$ at varying values of the Flory parameter $\chi$ (line thickness), from Eq. (31a); *Subplot:* $\bar{\psi}_m$ versus $\chi$ at varying $J$ (line thickness), from Eq. (31a).

Figure 3 plots the relations expressed in Eq. (30a) and (31a), with the normalized elastic free energy $\bar{\psi}_e/n$ on the left and the mixing free energy $\bar{\psi}_m$ on the right.

In Figure 3-left, $\bar{\psi}_e/n$ increases with an increased swelling ratio $J$ as the chains become more stretched, thus decreasing the entropy of the network. Decreasing the chain microstretch $\lambda_M$ leads to a greater $\bar{\psi}_e/n$, and here the difference between two red curves represents the injected energy in the fast motor (*FM*) activation case for CS-MM, where MMs instantaneously evolve the microstructure, thereby prompting contraction. The log-log plots in both the red curves and the blue curves, in the inset of this figure, show the trend

$$\frac{\bar{\psi}_e}{n} \sim \frac{3}{2} \frac{J^{2/3}}{\lambda_M^2} \tag{30b}$$

for large $J$ and/or small $\lambda_M$. In Figure 3-right, the mixing energy $\bar{\psi}_m$ decreases with increased $J$ as more solvent is added to the mixture, thus increasing entropy. The semi-log plot in the red curves shows the asymptotic saturation energy

$$\bar{\psi}_m \sim \chi - 1 \tag{31b}$$

for large $J$. This also explains the linear correlation between $\bar{\psi}_m$ and $\chi$ in the blue curves in the inset.

*Energetics of Contraction*

The conditions $\chi = -1.1$, $J_0 = 1000$, $n_0 = 1.61 \cdot 10^{-5}$, and $\lambda_{M,0} = 1$, highlighted in Figure 2 with blue circles, in this case, give $\bar{\psi}_{e,0}/n_0 = 1.42 \cdot 10^2$, $\bar{\psi}_{e,0} = 2.28 \cdot 10^{-3}$ and $\bar{\psi}_{m,0} = -2.1$. Figure 4 plots the mechanical work performed by MM in the fast activation regime (*FM*), $\bar{w}_{FM}$, and that in the slow activation regime, $\bar{w}_{SM}$, as a function of the contraction ratio $J_f/J_0$ and initial conditions given by $\chi = -1.1$, $J_0 = (10, 100, 1000)$, and the corresponding $n_0 = (4.5 \cdot 10^{-2}, 7.8 \cdot 10^{-4}, 1.6 \cdot 10^{-5})$. In this figure, on the left we have DC-MM, *i.e.* by evolving $n$, from an initial $n = n_0$, at steady $\lambda_M = 1$. On the right we have CS-MM, *i.e.* by evolving $\lambda_M$, from an initial $\lambda_M = 1$, at steady $n = n_0$. In this figure we can observe that the work required for contraction is proportional to the initial crosslink density $n_0$ and to the degree of contraction (high contraction giving low $J_f/J_i$). For $J_f/J_i \to 0$, we have that $\bar{w}_{SM}, \bar{w}_{FM} \to \infty$ (maximum contraction), while for $J_f/J_i = 1$ we have $\bar{w}_{SM} = \bar{w}_{FM} = 0$ (no contraction). These semi-log plots show a sharp increment in mechanical work for mild contractions ($J_f/J_i \sim 1$) as well as for maximal contractions ($J_f/J_i \to 0$), while intermediate contractions require more moderate increments in mechanical work. From the plots in this figure, we can observe that DC-MM and CS-MM require nearly identical $\bar{w}_{SM}$ and $\bar{w}_{FM}$ for a given $J_f/J_0$ and for higher initial swelling ratios, namely $J_0 = 100$, and 1000. We can, however, observe slightly higher $\bar{w}_{SM}$ for CS-MM at $J_0 = 100$ and very high contraction, *i.e.* $J_f/J_0 = 0.1$. For smaller initial swelling ratio, *i.e.* $J_0 = 10$, we have a significant distinction between DC-MM and CS-MM. Here, we can see that CS-MM have larger $\bar{w}_{SM}$ and smaller $\bar{w}_{FM}$ compared to DC-MM. It should be finally noted that the trends observed in Figure 4 depend on the constitutive relations used to describe chemical affinity and network stiffening, energetically, hence on the formulations of $\psi_e$ and $\psi_m$.

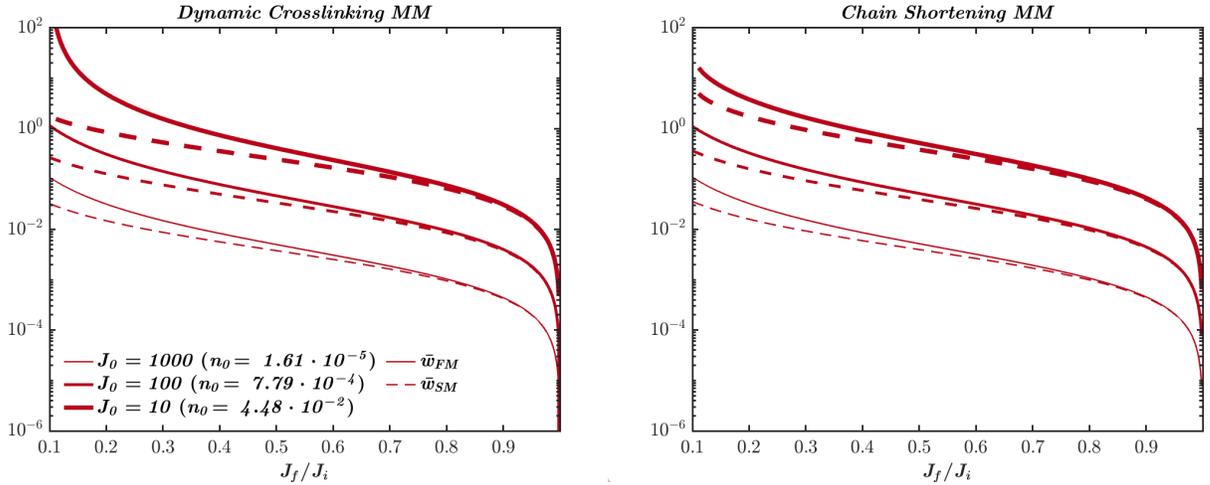

**Figure 4** – Dimensionless mechanical work, $\bar{w}_{FM}$ and $\bar{w}_{SM}$, performed by molecular motors as a function of the contraction ratio $J_f/J_0$ for both fast motor activation (*FM*, solid line) and slow motor activation (*SM*, dashed line). *Left*: Dynamic crosslinking molecular motors (DC-MM). *Right*: Chain-shortening molecular motors (CS-MM). Both left and right figures consider varying initial swelling ratio $J_0$ (varying line thickness), where the initial chain density, per unit solvent molecule, $n_0$ is calculated from the condition of equilibrium at Eq. (25), with $\chi = -1.1$.

Because

$$\bar{w}_{SM} = \int_{J_0}^{J_f} \frac{\partial \bar{\psi}}{\partial \zeta_c} \frac{\partial \zeta_c}{\partial J} dJ \qquad (32)$$

with $\zeta_c = n$ or $\lambda_M$, if we assume that both $J_0$ and $J_f$ are large, we can use Eq. (25b), (27b), and (30b) to obtain the estimation

$$\overline{w}_{SM} \sim \frac{5}{4}\frac{(1-2\chi)}{J_0}\left(\frac{1}{J_f/J_0} - 1\right) \tag{33}$$

Eq. (33) confirms that the SM work, $\overline{w}_{SM}$, is inversely proportional to the ratio $J_f/J_0$ and to the initial swelling ratio $J_0$, as observed in Figure 4.

*Experimental comparison*

The energetic requirement for contraction, predicted from our model, can be compared with experimental results from literature. Bendix et al. [23] conducted cytoskeletal gel contraction experiments observing the effects of chain density and motor concentration on contraction. They found that certain configurations yield observable contraction, while others yielded no observable contraction. Our model can explain this transition based on the required mechanical energy for contraction. The experiments consisted in varying crosslinker (α-actinin) and molecular motor (myosin II) concentrations, $N_\alpha$ and $N_{MM}$ respectively, with constant concentration of actin monomers, $N_A$. The gel specimens tested are of millimeter size, and the observable contractions were defined as achieving 90% volumetric deswelling ($J_f/J_i = 0.1$) in a timespan below 60 minutes. Given this long timescale of observation, related to specimen size, we consider contractile networks as those with sufficient concentration of MM so that they can exert the minimum work required for very slow contraction, *i.e.* $w_{SM}$. Bendix et al. [23] reports the fixed concentration of actin monomers, as well as the ratios of myosin molecular motors $R_{M:A} = N_{MM}/N_A$ and crosslinker α-actinin $R_{\alpha:A} = N_\alpha/N_A$ to actin. As explained in Appendix C, a portion of α-actinin molecules is required to form the actin filament bundles generating the polymer chain, thus unavailable to form chain-chain crosslinks, giving the ratio $R_{b:A} = N_b/N_A$. The crosslink (chain) density $N$ relates to the crosslinker concentration $N_\alpha$ as $N = N_\alpha - N_b$ [27-28], giving

$$N = (R_{\alpha:A} - R_{b:A})N_A \tag{34}$$

Eq. (34) assumes that, after bundling, each new crosslinking molecule splits an existing chain into two, therefore adding a new chain to the network.

We assume that the work provided by the MMs scales with MM concentration as $w_{SM} = w_M N_{MM}$, giving

$$w_{SM} = w_M R_{M:A} N_A \tag{35}$$

where $w_M$ is the work provided by one MM. Eq (35) is based on the hypothesis of unlimited ATP availability and that all motors equally contribute to provide a fixed mechanical work [29].

To determine the parameters of the gel at the equilibrium swollen state (0), for each experimental value of $R_{\alpha:A}$, we estimate $J_0$ from the chemo-mechanical equilibrium in Eq. (25) with $\lambda_M = 1$ and $n_0 = N_0 \Omega$ from Eq. (34) ($N = N_0$). Here, $N_A$ is extracted from the *current* density of actin monomers $\widehat{N}_A = N_A/J_0$. Note that, $\chi$ is undefined, and we extract it from a test case in [23] as explained in Appendix C, where $J_0 = 1000$ and $n_0 = 1.61 \cdot 10^{-5}$, giving $\chi = -1.1$. The latter is a material parameter, and thus is constant in each experiment. To determine the parameters of the gel at the final state (*f*), we enforce again chemo-mechanical equilibrium from Eq. (25). From this equation, with $J_f = 0.1 J_0$ and $\chi$, we can extract $n = n_f$, under $\lambda_M = 1$ for DC-MM or $\lambda_M$ from $n = n_0$ for CS-MM. Finally, by substituting Eq. (30) into (24), we can calculate the work $\overline{w}_{SM} = w_{SM}\Omega/kT$. This result can then be substituted into Eq. (35) to extract the minimum required MM density $N_M$, based on the value of $w_M$, which is here taken as a fitting parameter.

Figure 5 shows a comparison between our estimated minimum MM density required to contraction (with red shading indicating non-contractile regions) with the observed contractile (black circles) and non-contractile (red crosses) conditions in [23]. Our estimation is shown in the bottom solid line for CS-MM, and dashed line for DC-MM. For most of the experimental data points, except some outliers, our estimate agrees with experimental evidence. We speculate that some outlier data points, where contractile gels appear as non-contractile, or vice-versa, are due to geometrical imperfections. In this comparison we have adopted $w_M = 1.9 \cdot 10^7 \, kT$ for both CS-MM and DC-MM. These values are equivalent to approximately 1.3 ATP molecules consumed per myosin head per second in the timespan of one hour, as detailed in Appendix C. Experimental observations reported 0.5-5 ATP molecules consumed per myosin head per second [30]. Because $w_M$ estimates the effective work produced by the MM, the real ATP energy consumed is expected to be higher. The MM efficiency in this case can be estimated from a minimum of $1.3/5 = 26\%$ to a maximum of $1.3/0.5 = 86\%$. The upper bound in MM concentration in Figure 5 is due to a maximum strain energy per chain in the final state, $\psi_{e,f}/N_0$, above which the α-actinin crosslink breaks. In these conditions, contraction is prevented. We adopt this energy maximum as $\psi_{e,max}/N_0 = 2.44 \cdot 10^3 \, kT$ to fit the upper bound of the contractile region for both CS-MM and DC-MM. This value corresponds to the energy required to break one crosslink bond between the chains, and is equivalent to the energy of approximately 600 hydrogen bonds (see Appendix C). In regions of high crosslinking, such as on the right side of Figure 5, contraction can become much slower, as discussed by [22]. In the experimental timeframe observed by [23], contractions requiring more than 60 minutes were ignored, therefore classifying such trials as non-contractile. We speculate that the networks associated with the red crosses on the right side of Figure 5 may still be contractile, as predicted by our model, but requiring more than 60 minutes.

The contractile region predicted by our model (white shading) agrees with experiments for the lower and upper bounds in Figure 5, highlighting the energetic limit of contraction for cytoskeletal network and the structural limit of the actin backbone, respectively. Both CS-MM and DC-MM give similar predictions, with CS-MM being slightly more accurate.

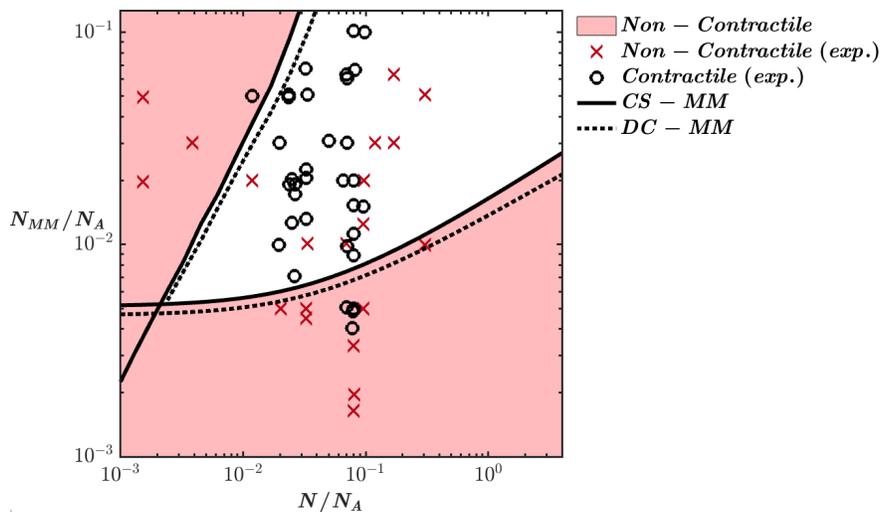

**Figure 5** – Contractility map evidencing the contraction requirements for a cytoskeletal gel having chain density $N$ and molecular motor density $N_{MM}$. Both $N$ and $N_{MM}$ are normalized by the density of actin monomers $N_A$. The model prediction is plotted with black lines (solid for CS-MM, and dashed for DC-MM), while the experimental data points, from [23] (*Contractile exp.*), are reported with red crosses, to indicate non-contractile gels, and black circles, to indicate contractile gels. The minimum energetic requirement from the model prediction is taken from $w_{SM}$ (Figure 4) to provide the lower limit, (bottom black lines). The limit in elastic strain energy stored in each chain prior to chain rupture, $\psi_{e,max}/N$, provides the upper limit (top black lines). Red shading indicates non-contractile regions, violating either the lower or the upper limits.

It should be noted that our model considers the polymer backbone to be composed of Gaussian (slack) chains. This hypothesis limits the reliability of our model, particularly in highly crosslinked networks (right side of Figure 5). Moreover, in this investigation we assumed, for simplicity, that CS-MM and DC-MM act as separate mechanisms. In general, both mechanisms can occur at the same time in a gel, and one should study their coupled effects during contraction. Our simplified approach provides useful results if one assumes that one mechanism prevails over the other. In actin gels equipped with myosin motors, such as the one used in [23], chain shortening is commonly the predominant mechanism. This is because actin polymerization and actin-myosin binding, which provide dynamic crosslinking, occur initially and leave the gel in a stiffened state, while myosin progressively shortens actin chains. This conclusion might explain why our prediction from CS-MM (solid lines), in Figure 5, yields slightly better agreement with experiments.

**Conclusion**

Contractile cytoskeletal networks exhibit a range of energetically motivated contractions caused by molecular motor activity. Our model introduces a more physical representation of CS-MM activation, where, from a mechanical standpoint, the effective chain length (composed of only engaged monomers) shortens according to the newly introduced 'microstretch' parameter. This approach yields similar results to that of previously employed chain density evolution models, particularly for highly swollen gels. Via simple steady state energetic analysis, under the simplest case of isotropic contraction, we quantify the mechanical energy required for contraction as a function of polymer chain density and molecular motor density. We identify two limit cases, (*FM*) fast molecular motor (MM) activation for which MMs provide all the available mechanical energy immediately and (*SM*) slow MM activation where the timescale is much longer, and contraction is much slower. These two cases represent the maximum and minimum timescales for the contraction process. They also represent the two limits in the efficiency of energy transduction, where *sm* gives the highest efficiency and *FM* gives the lowest efficiency. This is because, to achieve the same amount of contraction, the mechanical work required is highest with fast MM activation ($w_{FM}$) and lowest with slow MM activation ($w_{sm}$), *i.e.* $w_{SM} < w < w_{FM}$. We observe that the energetic cost of contraction fulfilled by MM is proportional to chain density and MM density. Finally, we compare our results with experiments and observe good agreement, where the 90% contraction boundary is predicted by our model.

Our model provides an accurate and simple description of the energetic landscape of contraction for gels composed of slack chains. It also provides the minimum requirement of actin and myosin density to allow cytoskeletal contraction in such systems. Further development of this model will allow for a more comprehensive description of the cytoskeleton, for example one composed of tight chains.

**Table 1** – Model parameters with their symbol notation, description, nominal value and source

| Symbol | Description | Nominal Value | Source |
|---|---|---|---|
| $kT$ | Unit energy, $k$ Boltzmann constant, $T$ absolute temperature | $4.14 \cdot 10^{-21} \, J$ | [12] |
| $\Omega$ | Molecular volume of the solvent | $1.41 \cdot 10^{-2} \, nm^3$ | [12] |
| $J_0$ | (Initial) swelling ratio of the gel | 1000 | [23] |
| $n_0 = \Omega N_0$ | (Initial) dimensionless crosslink density | $1.63 \cdot 10^{-5}$ | [23] |
| $\chi$ | Flory parameter | $-1.1$ | Appendix C |
| $N_A$ | Actin monomers concentration per unit reference (dry polymer) volume | $1.43 \cdot 10^{25} \, m^{-3}$ | [23], Appendix C |
| $R_{\alpha:A} = N_\alpha/N_A$ | Ratio of α-actinin crosslinker concentration ($N_\alpha$) to actin monomer concentration ($N_A$) | 0.11 | [23], Appendix C |
| $R_{b:A} = N_b/N_A$ | Ratio of bundling α-actinin concentration ($N_b$) to actin concentration ($N_A$) | 0.03 | [23], Appendix C |
| $w_M$ | Total work provided by one molecular motor (MM) after one hour of activity | $1.9 \cdot 10^7 \, kT$ | Appendix C |
| $\psi_{e,max}/N$ | Maximum elastic strain energy stored by a chain prior to its rupture | $2.44 \cdot 10^3 \, kT$ | Appendix C |
| $V$ | Equilibrium volume of the passive gel | $8.8 \cdot 10^{-10} \, m^3$ | Appendix C |
| $V_p$ | Volume of polymer chains | $8.24 \cdot 10^{-13} \, m^3$ | Appendix C |
| $d_A$ | Actin monomer diameter | $5 \, nm$ | [31] |
| $J_f/J_i$ | Volumetric contraction (deswelling) ratio | 0.1 | [23] |

## Acknowledgments

This work was supported by the New Frontiers in Research Funds – Exploration (NFRFE-2018-00730) and by the Natural Sciences and Engineering Research Council of Canada (NSERC) (RGPIN-2017-04464).

## Authors contributions

Conceptualization: M.F., M.B.; Methodology/Software: M.F., A.K., M.H., M.B.; Validation: M.F.; Writing: M.F., A.K., G.J.E, D.A., M.B.

## Appendix A

The first law of thermodynamics gives

$$\frac{\mathrm{d}}{\mathrm{d}t} \int_{V_d} e \, \mathrm{d}V_d = \int_{S_d} T_j \dot{x}_j \mathrm{d}S_d + \int_{V_d} B_j \dot{x}_j \mathrm{d}V_d - \int_{S_d} N_j J_j^h \mathrm{d}S_d - \int_{S_d} h^q N_j J_j^q \mathrm{d}S_d \tag{A1}$$

where repeated indices indicate a sum. Here, $e$ is the internal energy per unit reference volume, $J_j^h$ is the heat flux in the reference configuration, $\dot{x}_j$ is the velocity of elements in the volume and thus is the rate of change of $x_j$ ( $\dot{}$ implies time derivative), $h^q$ is the partial molar enthalpy of species $q$, $B_j$ the body force vector, $N_j$ the unit normal to boundary surfaces, and $T_j$ the external traction force. Use of the divergence theorem and the principle of virtual powers [8] yields

$$\dot{e} = t_{jk}\dot{F}_{jk} - \frac{\partial}{\partial x_j}\left(J_j^h + h^q J_j^q\right) \tag{A2}$$

Here, the first term on the right-hand side is the mechanical power from external forces, while the second term is the energy loss, per unit time and reference volume and unit time, due to heat and mass exchange.

Entropy conservation [8] gives

$$\frac{d}{dt}\int_{V_d}\eta\,dV_d = \int_{V_d}\dot{\eta}^P dV_d - \int_{S_d} N_j \frac{J_j^h}{T} dS_d - \int_{S_d}\eta^q N_j J_j^q dS_d \tag{A3}$$

where $\eta$ is the entropy per unit volume in the reference state, $\eta^q$ is the partial molar entropy of species $q$ and $\dot{\eta}^P$ is the rate of entropy production per unit volume in the reference state. With divergence theorem, Eq. (A3) leads to

$$\dot{\eta} = \dot{\eta}^P - \frac{\partial}{\partial x_j}\left(\frac{J_j^h}{T} + \eta^q J_j^q\right) \tag{A4}$$

Here, the first term on the right-hand side is the entropy generation, while the second term is entropy loss, per unit time and reference volume, due to heat and mass exchange.

The Helmholtz energy per unit reference volume is

$$\psi = e - T\eta \tag{A5}$$

From Eq. (A2), (A4), and (A5), the rate of change of the Helmholtz free energy is

$$\dot{\psi} = t_{jk}\dot{F}_{jk} - \dot{\eta}^P T - \eta\dot{T} - \frac{J_j^h}{T}\frac{\partial T}{\partial x_j} - \frac{\partial}{\partial x_j}\left(h^q J_j^q\right) + T\frac{\partial}{\partial x_j}\left(\eta^q J_j^q\right) \tag{A6}$$

Here, the first term on the right-hand side is the mechanical power, per unit reference volume, done by external forces, and the second and third ones are free energy losses, per unit time and reference volume, due to entropy generation and temperature increment, respectively. The last three terms are free energy losses due to heat and mass exchange.

Next, we introduce $\mu^q$, the chemical potential of species $q$ as

$$\mu^q = h^q - T\eta^q \tag{A7}$$

and mass conservation

$$\dot{C}^q = Q^q - \frac{\partial J_j^q}{\partial x_j} \tag{A8}$$

with $Q^q$ the generation of specie $q$ due to chemical reactions.

Substitution of Eq. (A7) and (A8) into (A6) gives

$$\dot{\psi} = t_{jk}\dot{F}_{jk} - \dot{\eta}^p T - \eta\dot{T} - \frac{J_j^h}{T}\frac{\partial T}{\partial x_j} - \left(\frac{\partial \mu^q}{\partial x_j} + \eta^q\frac{\partial T}{\partial x_j}\right)J_j^q + \mu^q(\dot{C}^q - Q^q) \tag{A9}$$

Assuming the Helmholtz free energy has the functional dependence $\psi = \psi(T, \boldsymbol{F}, \boldsymbol{\zeta}, \boldsymbol{C})$, as described in (1), with (A9) we solve for the entropy production as

$$\dot{\eta}^p T = \left(t_{jk} - \frac{\partial \psi}{\partial F_{jk}}\right)\dot{F}_{jk} + \left(\mu^q - \frac{\partial \psi}{\partial C_q}\right)\dot{C}_q - \left(\eta + \frac{\partial \psi}{\partial T}\right)\dot{T} - \frac{J_j^h}{T}\frac{\partial T}{\partial x_j} - \left(\frac{\partial \mu^q}{\partial x_j} + \eta^q\frac{\partial T}{\partial x_j}\right)J_j^q - \mu^q Q^q - \frac{\partial \psi}{\partial \zeta_c}\dot{\zeta}_c \tag{A10}$$

From the second law of thermodynamics, the rate of entropy production must be $\dot{\eta}^p \geq 0$. Considering chemical equilibrium, homogeneous and stationary distribution of temperature, and species conservation, which also implies no flux of species, and no microstructural change, the first 3 terms on the right-hand side of Eq. (A10) are $\geq 0$ for all deformation rates, concentration changes, and temperature adjustments either positive or negative. These considerations provide equations (4) and (5) and

$$\frac{\partial \psi}{\partial T} = -\eta \tag{A11}$$

In Eq. (5), the specie considered is water, and the index $q$ is omitted.

With the above satisfied, we are left with the remaining terms in Eq. (A10), thus

$$-\frac{J_j^h}{T}\frac{\partial T}{\partial x_j} - \left(\frac{\partial \mu^q}{\partial x_j} + \eta^q\frac{\partial T}{\partial x_j}\right)J_j^q - \mu^q Q^q - \frac{\partial \psi}{\partial \zeta_c}\dot{\zeta}_c \geq 0 \tag{A12}$$

The first term on the left-hand side of Eq. (A12) governs the heat flow, while the second term governs the transport of chemical species. The last two terms govern the transient microstructural processes, where the energy production from chemical reactions can produce a microstructural change [8]. Considering that $Q^q = 0$ for all species except ATP and ADP (adenosine triphosphate and diphosphate, respectively), we have that

$$(\mu^{ATP} - \mu^{ADP})Q^h \geq \frac{\partial \psi}{\partial \zeta_c}\dot{\zeta}_c \tag{A13}$$

with $Q^h$ the rate of consumption (production) of ATP (ADP) during hydrolysis. The inequality holds so long that the efficiency of the mechanotransduction process is less than unity. We can thus rewrite Eq. (A13) as

$$\frac{\partial \psi}{\partial \zeta_c}\dot{\zeta}_c = \dot{w}_c \tag{A14a}$$

where $\dot{w}_c$ is the mechanical power density, per unit reference volume, required to evolve the microstructural variable $\zeta_c$ at the rate $\dot{\zeta}_c$. If these mechanisms are driven by molecular motors (MMs) hydrolyzing ATP (adenosine triphosphate), we can write $\dot{w}_c = f_c \dot{w}$, with $f_c$ the portion of energy required to evolve $\zeta_c$, and $\dot{w} = \sum_c \dot{w}_c$ the total power generated by ATP hydrolysis. The latter is given by

$$\dot{w} = \varphi(\mu^{ATP} - \mu^{ADP})Q^h \tag{A14b}$$

where $\varphi$ is the energy transduction efficiency of the MMs. Eq. (A14) is rewritten in (6) in differential form.

# Appendix B

In this appendix we provide the mathematical derivation for the strain energy density incorporating chain shortening. The statistical configuration of the generic Gaussian chain $i$ is described by its end-to-end vector $\boldsymbol{R}_i = (X_i, Y_i, Z_i)$, and its probability density function (PDF) is

$$f(\boldsymbol{R}_i) = \left(\frac{3}{2\pi\langle R^2\rangle}\right)^{3/2} \exp\left(-\frac{3R_i^2}{2\langle R^2\rangle}\right) \tag{B1a}$$

where $R = \sqrt{X^2 + Y^2 + Z^2}$ is the end-to-end length, and

$$\langle R^2\rangle = \int_{-\infty}^{\infty}\int_{-\infty}^{\infty}\int_{-\infty}^{\infty} R^2 f(\boldsymbol{R}) dX\, dY\, dZ \tag{B1b}$$

is the mean squared end-to-end distance of the network. The latter is also given by

$$\langle R^2\rangle = m_0 b^2 \tag{B2}$$

where $m_0$ is the average number of monomers in each chain of the network, and $b$ the average length of each segment in the chain. Gaussian chains are assumed to be non-interacting, and the free energy $g(\boldsymbol{R}_i)$ of the generic $i$-th Gaussian chain is described by Boltzmann's entropy equation

$$g(\boldsymbol{R}_i) = -kT \ln[P(\boldsymbol{R}_i)] \tag{B3}$$

where $P(\boldsymbol{R}_i)$ is the probability for chain $i$ of having the end-to-end vector equal to $\boldsymbol{R}_i$. This is given by its PDF as

$$P(\boldsymbol{R}_i) = f(\boldsymbol{R}_i) dX dY dZ \tag{B4}$$

After deformation of the network, the end-to-end vector of chain $i$ evolves from $\boldsymbol{R}_i$ to $\boldsymbol{r}_i = (x_i, y_i, z_i)$. Considering an affine deformation, by which all chains follow the same deformation tensor $\boldsymbol{F}$, we have that $\boldsymbol{r}_i = \boldsymbol{F} \boldsymbol{R}_i$. Take now the principal stretches $\lambda_x, \lambda_y, \lambda_z$, so that $x_i = \lambda_x X_i$, $y_i = \lambda_y Y_i$, $z_i = \lambda_z Z_i$, and $J = \lambda_x \lambda_y \lambda_z$ with $dxdydz = J\, dXdYdZ$. The change in the free energy change of the chain, due to deformation of the network, from Eq. (B3) and (B4) is

$$\Delta g_i = -kT \ln\left[\frac{f(\boldsymbol{F}\,\boldsymbol{R}_i)}{f(\boldsymbol{R}_i)}\right] - kT \ln J \tag{B5}$$

By substituting Eq. (B1) into (B5) we have

$$\Delta g_i = \frac{3}{2}kT\left[X_i^2\left(\frac{\lambda_x^2}{\langle r^2\rangle} - \frac{1}{\langle R^2\rangle}\right) + Y_i^2\left(\frac{\lambda_y^2}{\langle r^2\rangle} - \frac{1}{\langle R^2\rangle}\right) + Z_i^2\left(\frac{\lambda_z^2}{\langle r^2\rangle} - \frac{1}{\langle R^2\rangle}\right)\right] + \frac{3}{2}kT \ln\left(\frac{\langle r^2\rangle}{\langle R^2\rangle}\right) - kT \ln J \tag{B6}$$

The total free energy density of the network is given by the sum of the free energy of all the $N$ chains in the material point, giving

$$\psi_e = \sum_{i=1}^{N} \Delta g_i \tag{B7}$$

Not that $\sum_{i=1}^{N} R_i^2 = \langle R^2\rangle$, and, since the network has no directional configuration in the undeformed state, we have $\sum_{i=1}^{N} X_i^2 = \sum_{i=1}^{N} Y_i^2 = \sum_{i=1}^{N} Z_i^2 = \langle R^2\rangle/3$. From this, by substituting Eq. (B6) into (B7) we have

$$\psi_e = \frac{NkT}{2}\left[\frac{\lambda_x^2 + \lambda_y^2 + \lambda_z^2}{\lambda_M^2} - 2\ln\left(\frac{J}{\lambda_M^3}\right) - 3\right] \tag{B8}$$

where we have considered $\langle r^2\rangle = m\, b^2$, with $\lambda_M^2 = m/m_0$. In this case $m$ represent the number of engaged monomers between crosslinks in the deformed state and $m_0$ that in the undeformed state. A common assumption is that the chain length is unchanged, and so the number of engaged monomers between crosslinks, giving $m = m_0$ and $\lambda_M = 1$. This recovers traditional neo-Hookean elasticity. However, chain shortening molecular motors can reduce the effective number of monomers between crosslinks, leading to $m < m_0$, from which $\lambda_M < 1$. By substituting the invariant of the strain tensor $\lambda_x^2 + \lambda_y^2 + \lambda_z^2 = F_{ij}F_{ij}$ into Eq. (B8) we finally obtain (11).

**Appendix C**

The parameters of the model are based on experiments performed on actin gels by Bendix et al. [19], with data extracted from the supplementary material. The gel volume in the equilibrium swollen state is $V = 8.8 \cdot 10^{-10} \, m^3$ [23]. We consider an actin monomer as a spheroidal molecule of diameter $d_A = 5 \, nm$ [31]. Given the concentration of actin monomers in the current state (number of monomers per unit current volume) $\widehat{N}_A = 1.43 \cdot 10^{22} \, m^{-3}$, we can calculate the volume of the polymer as $V_p = \widehat{N}_A V_0 \pi d_A^3 / 6$, giving $V_p = 8.24 \cdot 10^{-13} \, m^3$. Because the equilibrium swelling ratio is $J_0 = V/V_p$, we have then $J_0 = 1068.45$, which we round to $J_0 \simeq 1000$. This gives the actin density in the reference state $N_A = \widehat{N}_A J_0 = 1.43 \cdot 10^{25} \, m^{-3}$. The cytoskeletal network is composed of chains, which are made of bundles of actin filaments. In each filament, on average, every 30 actin monomers, there is an α-actinin molecule [32]. This gives the ratio $R_{b:A} = 0.03$ in Eq. (31). For the example case of $R_{\alpha:A} = 0.11$ we have the crosslink density $N_0 = 1.14 \cdot 10^{24} \, m^{-3}$. Taking the diameter of a water molecule as $d_w = 0.3 \, nm$, we have then $\Omega = \pi d_w^3 / 6$, giving $\Omega \simeq 1.41 \cdot 10^{-29} \, m^3$. The nominal crosslink density becomes then $n_0 = N_0 \Omega = 1.61 \cdot 10^{-5}$. Eq. (25), with $\lambda_m = 1$, is satisfied and gives our initial conditions if $\chi = -1.1$.

Take $kT = 4.14 \cdot 10^{-21} \, J$ as a unit energy [12]. The total mechanical work provided by a single MM during the whole contraction process, $w_M$, is used in our investigation as a fitting parameter to predict the lower bound estimation of the minimum required MM density for contraction in Figure 5. This, for both CS-MM and DC-MM, this computes to $w_M = 8 \cdot 10^{-14} \, J = 1.9 \cdot 10^7 \, kT$. Taking the energy of ATP hydrolysis as $12 \, kT$ per ATP molecule [31], this computes to the equivalent energy provided by approximately 417 ATP molecules hydrolyzing on each motor every second for a period of one hour. Considering each myosin motor has roughly 300 heads [31], this computes to the equivalent of roughly 1.3 ATP molecules consumer per head every second for one hour. It should, however, be noted that the efficiency of mechanotransduction is less than unity, thus the real ATP consumption is expected to be higher than the above estimation. The maximum strain energy stored by one polymer chain prior to crosslink (chain) rupture is used as a fitting parameter for the upper bound estimation of the required MM density in Figure 5. This computes to $\psi_{e,max}/N_0 = 1.01 \cdot 10^{-16} \, J = 2.44 \cdot 10^3 \, kT$ for both CS-MM and DC-MM. Given a breaking a hydrogen bond requires approximately $10 \, kJ/mol = 4 \, kT$ [33], this strain energy is equivalent to breaking approximately 600 hydrogen bonds.